\documentclass[journal=jpclcd,manuscript=letter]{achemso}

\usepackage[version=3]{mhchem} 
\usepackage{graphicx}
\usepackage{amsmath}
\usepackage{amssymb}
\usepackage{color}

\author{Ivan A. Pshenichnyuk}
\email{Ivan.Pshenichnyuk@physik.uni-erlangen.de}
\author{Pedro B. Coto}
\author{Susanne Leitherer}
\author{Michael Thoss}
\email{Michael.Thoss@physik.uni-erlangen.de}
\affiliation[FAU Erlangen-N\"{u}rnberg]
{Institute of Theoretical Physics and Interdisciplinary Center for Molecular
  Materials,  Friedrich-Alexander-Universit\"{a}t Erlangen-N\"{u}rnberg,
  Staudtstrasse 7/B2, D-91058 Erlangen, Germany}
\title[Charge Transport in Pentacene-Graphene Nanojunctions]
{Charge Transport in Pentacene-Graphene Nanojunctions}


\begin{document}

\begin{abstract}
We investigate charge transport in pentacene-graphene nanojunctions employing density functional theory (DFT) electronic structure calculations and the Landauer transport formalism.
The results show that the unique electronic properties of graphene strongly influence the transport in the nanojunctions. 
In particular, edge states in graphene electrodes with zigzag termination result in additional transport channels close to the Fermi energy which deeply affects the conductance at small
bias voltages.
Investigating different linker groups as well as chemical substitution, we demonstrate how the transport properties are furthermore influenced by the molecule-lead coupling and the energy level lineup.
\end{abstract}

Molecular junctions, consisting of single molecules that are chemically bound
to electrodes, represent interesting systems to investigate and understand
mechanisms of nonequilibrium transport processes at the nanoscale.
Recent experimental\cite{reed-1997,park-2000,smit-2002,reichert-2002,qiu-2004,elbing-2005,tao-2006,ioffe-2008,vandermolen-2010,ballmann-2012}
and theoretical \cite{hanggi-2002,nitzan-2003,cuniberti,joachim-2005,galperin-2008,cuevas,hartle-2011} works
have elucidated the conductance properties of molecular junctions
and revealed a wealth of interesting transport phenomena 
such as Coulomb blockade, Kondo effects, negative differential resistance, vibronic effects and
local heating, as well as switching and hysteresis.

Most experiments on single-molecule conductance have employed metal
electrodes. Carbon-based materials, such as carbon nanotubes or graphene,
represent another promising class of materials for electrodes. 
Studies of molecular contacts with carbon nanotubes have demonstrated
a number of advantages 
\cite{feldman-2008,delvalle-2007} as compared
to metal electrodes, in particular, rigidity, mechanical stability and
thus  a more precise control of the
molecule-electrode contact geometries.
Graphene, furthermore offers excellent electronic
properties \cite{castroneto-2009}, especially high electron mobility and, furthermore, may
facilitate optical addressability of the nanocontact for optoelectronic
applications\cite{liu-2009}. Electron transport through  molecular films connecting graphene
electrodes \cite{cao-2009} and molecular
junctions with few layer graphene electrodes
\cite{prins-2011} were recently reported.
Furthermore, using electron beam lithography, all-carbon junctions consisting
of a carbon chain between graphene electrodes were realized \cite{jin-2009}.

In this work, we study the conductance properties of molecule-graphene
nanojunctions employing a combination of density functional theory (DFT) 
calculations and the Landauer transport formalism. 
Considering several different examples, 
we show that charge transport in molecule-graphene nanojunctions is strongly
influenced by the edge structure of the graphene electrodes. 
In particular, for graphene
electrodes with zigzag termination, edge states may result in additional
transport channels close to the Fermi energy that affect the conductance at
small voltages profoundly. These findings extend previous theoretical work on
graphene-molecular  junctions \cite{bergvall-2011, ryndyk-2012, zanolli-2010}.

\begin{figure}
  \includegraphics{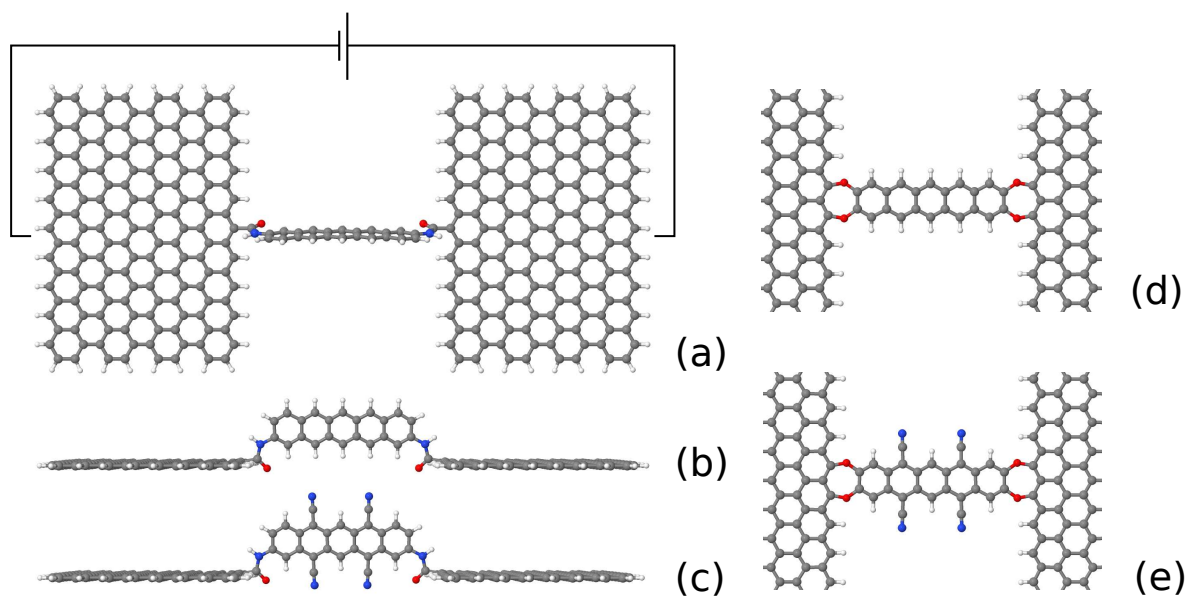}
  \caption{Molecule-graphene junctions with a covalently bound
      pentacene bridge  investigated in this work. (a) Overall
      structure  of a junction. Two
      types of  molecular bridges without (b,d) and with (c,e) CN-groups are
      considered.  Two different geometrical arrangements of the molecular
      bridge have  been analyzed: twisted geometries (b,c), where the molecular
      bridge is  weakly coupled to the leads through amide-binding groups
      and nearly flat geometries (d,e), where the bridge is strongly coupled
      via oxygen atoms to the leads.}
  \label{f1}
\end{figure}

As prototypical  systems, we consider molecular bridges based on pentacene
  as the central  unit (see Fig.~\ref{f1}). The extended $\pi$-electron system of pentacene facilitates electron 
transport and makes this system, in combination with its optical properties, a
promising material for molecular devices. Pentacene-based systems have been
used, for example, in thin-film organic electronic devices \cite{forrest-2004,katz-2009}. 
In the graphene-molecule junctions depicted in Fig.\ref{f1}b and c, 
the  molecular bridge is bound covalently via amide linker groups to
two graphene nanocontacts. This binding group has been used successfully for molecular
nanojunctions with carbon nanotubes \cite{feldman-2008}. The coupling of the 
molecule-graphene nanojunction to leads and an external
circuit are modeled by self energies and absorbing potentials (\textit{vide infra}).
To analyze the influence of graphene edge states on the conduction mechanism,
we investigate both zigzag-terminated graphene nanocontacts, as
depicted in Fig.\ref{f1}, and armchair termination.

In addition to the basic pentacene molecular bridge, we consider
CN-substituted molecular bridges (Fig.\ref{f1}c and e). The electron-withdrawing
CN-group allows one to shift the location of the molecular energy levels relative to
the Fermi energy of graphene and thus to vary the conductance at small voltages. 
Due to the almost perpendicular geometry of the
molecular bridge with respect to graphene, the molecular junctions depicted in
Figs.\ref{f1}b and c exhibit only weak coupling between the electronic $\pi$-orbitals
of pentacene and graphene and thus low conduction. 
As an example for strongly coupled systems, we consider the
systems depicted in Fig.\ref{f1}d and e,  where the binding to the
nanocontacts is achieved via oxygen atoms forming oxepine rings resulting in
an overall planar geometry and a strong electronic interaction between the
central pentacene molecule and the graphene nanocontacts. 

The theoretical methodology used to study charge transport in the systems considered
is based on a combination of DFT calculations
to characterize the geometry and electronic structure
of the systems and Landauer theory to simulate the transport properties. 
Closed shell
DFT calculations using the B3LYP exchange correlation functional together with
the  6-31G(d) basis set are used to model the different systems investigated in
this work. To avoid dangling bonds, the graphene edges have been saturated
using  hydrogen atoms. During the optimization process, geometrical
constraints  have been imposed to prevent artificial distortions of the
nanocontacts. All  calculations were carried out using G09\cite{G09} and
TURBOMOLE\cite{TURBOMOLE} codes.
 Following previous work,\cite{benesch-2008} the Kohn-Sham matrix obtained for
 the relaxed junction is identified with the Hamiltonian of the system.
To facilitate the transport calculations, the Hamiltonian matrix
is partitioned into blocks that correspond
to  the molecular bridge $H_m$, left and right contacts $H_l$ and $H_r$ and
coupling  between them $V_{lm}$, $V_{ml}$, $V_{mr}$ and $V_{rm}$,
respectively, as described by \citet{benesch-2008}. 
Direct coupling  between the left and the right contacts is
neglected. To model the effect that extended graphene
  contacts have in the electron transport,
we employ absorbing boundary conditions described by complex absorbing
potentials $-iW_{\alpha}$ ($\alpha = l,r$), which are added to the Hamiltonian 
in the graphene contacts. Thus, the overall Hamiltonian $\tilde{H}$ reads
\begin{equation}
  \tilde{H} =
  \begin{pmatrix}
       H_l-iW_l & V_{lm} & 0       \\
       V_{ml}  & H_m    & V_{mr}  \\
       0       & V_{rm} & H_r-iW_r \\
  \end{pmatrix}.
  \label{ham}
\end{equation}
The matrices describing the  absorbing potentials  are assumed  
to be diagonal, $(W_{\alpha})_{ij} = w_{\alpha i}\delta_{ij}$. 
Following previous work\cite{kopf-2004}, a polynomial function was employed to
model $w_{\alpha i}$. Accordingly, 
the elements of the matrices, which
correspond  to the $i$th atom of the nanocontact, are described by the formula
\begin{equation}\label{ap}
  w_{\alpha i} = a|x_i-x_{\alpha}^0|^4,
\end{equation}
where $x_i$ is the $x$-coordinate of the atom, assuming that the $x$ axis  is
directed along the molecular bridge, and $x_{\alpha}^0$ is selected to be close to
the  point where the molecular bridge is connected to the nanocontact.
To determine the empirical parameter $a$, test calculations over a wide range of values were carried out.
On the basis of that, the value of $a$ was chosen from a stable parameter range, where the results do not depend on $a$ \citet{kopf-2004}.

The transport  properties of the nanojunctions are described by
the transmission function. On the basis of the model introduced above, the
transmission function for an electron with energy $E$
is given by the expression \cite{datta}
\begin{equation}\label{tr}
  t(E) = \text{tr} \{ \Gamma_r G_m \Gamma_l G_m^{\dagger} \},
\end{equation}
where the trace is taken over the molecular degrees of freedom.
The transmission function close to the Fermi energy $E_f$
determines the conductance $\mathcal{G} =\frac{e^2}{\pi\hbar}t(E_f)$ at small bias voltages.

In Eq.\ (\ref{tr}),  $G_m$ denotes the Green's function projected on the molecular bridge
\begin{equation}\label{gf}
  G_m(E) = [E - H_m - \Sigma_l(E) - \Sigma_r(E) ]^{-1}.
\end{equation}
The matrix $\Gamma_{\alpha} \equiv -2\text{Im}\{\Sigma_{\alpha}\}$, which
characterizes the energy-dependent width of the molecular resonance states, is proportional
to the imaginary part of the self energy $\Sigma_{\alpha}$.
The self energy $\Sigma_{\alpha}(E)$ describes the coupling of the
molecular bridge to the graphene leads and is given by the expression
\begin{equation}\label{se}
\Sigma_{\alpha}(E) = V_{m\alpha} [E-H_{\alpha} + iW_{\alpha}]^{-1} V_{\alpha{m}}, \,\,\,\,\,\, \alpha = l,r.
\end{equation}

\begin{figure}
  \includegraphics{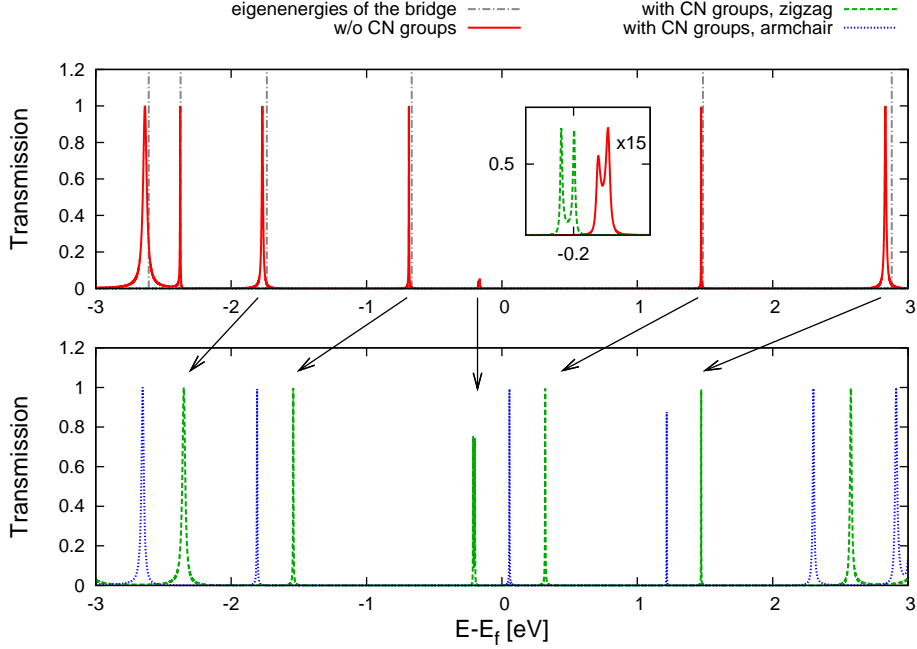}
  \caption{Transmission through a weakly coupled pentacene-graphene junction.
    The upper panel depicts the transmission function for the
    pentacene  bridge (solid line). The
    eigenenergies  of the bridge (dash-dotted line) are also shown. The lower panel
    demonstrates the  influence of CN groups on the transmission. The differences between
    zigzag  nanocontacts (dashed lines) and armchair nanocontacts (dotted
    line) are also shown. The inset highlights the edge induced peak with and
    without CN substitution.}
  \label{f2}
\end{figure}

\begin{figure}
  \includegraphics{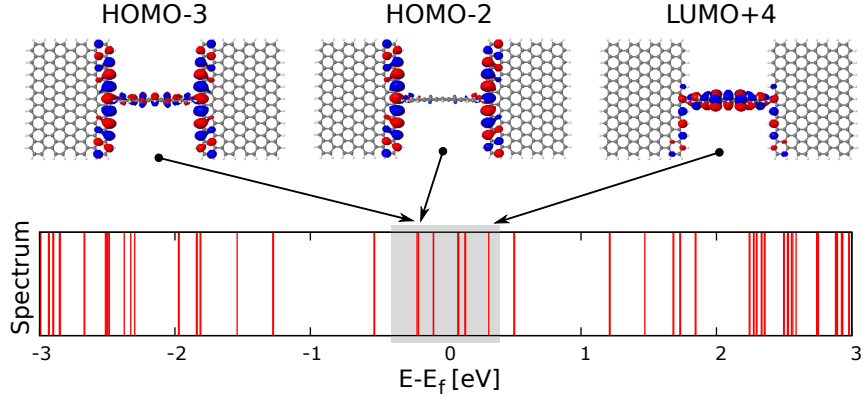}
  \caption{Energy spectrum of the molecular junction depicted in Fig.\ref{f1}c. The
    highlighted  area contains 8 edge states (including HOMO-3 and HOMO-4) and
    one  molecular state (LUMO+4), see text for details. Through
    CN-functionalization  of the pentacene bridge, the molecular state
    (LUMO+4) is  embedded into the family of edge states. The energies of the
    states  depicted correspond to peaks at $E=-0.2$ eV and $E=+0.2$ eV in
    Fig.\ref{f2}  (bottom, dashed line). }
  \label{f3}
\end{figure}

We first consider the amide-linked nanojunction with the
basic pentacene bridge (see Fig.\ref{f1}b).
Fig. \ref{f2} (top) shows the transmission function of this junction 
in the range of energies close to the Fermi level.
The transmission function exhibits a series of peak structures, which can be
classified into two types: 
(i) isolated resonance peaks with a maximum transmission value of unity, in
the following referred to as regular or molecular resonances, and
(ii) a small double peak structure close to the Fermi energy (see also the
inset in Fig.\ref{f2}). The former
are associated with resonant electron transport through the discrete energy levels
of the molecular bridge and thus appear at energies close to the
eigenenergies of the molecular levels. The width of these peaks varies
significantly, depending on the coupling of the specific molecular level to
the graphene leads. The latter double peak structure close
to the Fermi energy, on the other hand,
cannot be related directly to electronic states of the molecular bridge.
Furthermore, this peak structure does not correspond
to  direct tunneling between the graphene nanocontacts 
because this mechanism was explicitly  excluded 
in our treatment (cf.\ Eq.\ \ref{ham}). 
As discussed in detail below, we assign the double peak
  structure close  to the
Fermi energy  to an  additional transport channel caused by the edge states of the
graphene  nanocontacts (in the following referred to as edge induced transmission
channel). It is worthwhile to mention that similar surface
induced transport channels have also been found in carbon-nanotube-based
molecular  junctions \cite{gutierrez-2003, fagas-2004} as well as
in  molecular junctions with metal electrodes \cite{benesch-2006, xue-2003},
where  they are known as a metal-induced gap states. Due to the proximity of
the peak structure to the Fermi energy, it dominates the conductance at small bias voltages.

In the nanojunction considered in the top panel of Fig.\ref{f2}, 
the amplitude of the edge-induced transmission peak is rather small.
The amplitude  can be strongly
enhanced if molecular levels exist close to the Fermi energy.
This is demonstrated in the bottom
panel of Fig.\ref{f2}, which shows the transmission function for the 
CN-substituted pentacene bridge.  
The CN-substitution causes an overall shift of the molecular levels
to smaller energies. As a consequence, the energy level
associated  with the LUMO of
pentacene approaches the Fermi energy. As a result, the edge-induced transmission peaks
acquire additional intensity that increases their maximal value  to
almost unity. Furthermore, the interaction
between the molecular resonance state and the edge-induced transmission
peaks results in a splitting of the double peak.

To further analyze the edge-induced transmission channel, we
have  investigated
the electronic states of the overall nanojunction, as represented by their
molecular orbitals,  in a range of energies close to the
Fermi level  (see Fig.\ref{f3}). 
These states can be classified in two groups: (a) 
states localized preferentially in the graphene contacts and (b) molecular states
mostly localized on the bridge (see, e.g.\ LUMO+4 in Fig.\ref{f3}). A detailed
analysis of the (a)-type states shows that many of them are localized at the edge
of the  graphene contacts (see e.g.\ HOMO-3, HOMO-2 in Fig.\ref{f3}). These
states are  closely related to the well-known edge states of zigzag-terminated
graphene nanoribbons \cite{castroneto-2009}. Due to their
localization at  the molecule-graphene interface, they interact strongly with
the  molecular states and induce the additional transport channels. This can
be seen, for example,  in the spectrum depicted in Fig.\ref{f3} bottom, where the near
degenerate  pair of states HOMO-2 and HOMO-3 correspond to the induced
transmission peak  of the CN-substituted pentacene at $E=-0.2$ eV (see
Fig. \ref{f2}).  On the other hand, the molecular state LUMO+4 in Fig.\ref{f3}
represents  a regular transmission channel and corresponds to the peak at
$E=+0.2$ eV  (see Fig. \ref{f2}).

To unambiguously show that the edge-induced transmission channel appears as a
consequence of the existence of edge states in zigzag-terminated graphene
nanocontacts,  we have compared these results with the ones obtained for
molecular  junctions based on armchair-terminated graphene nanocontacts. As is
well-known, armchair edges of graphene do not possess
edge states\cite{wakabayashi-2009}. As a consequence, the transmission
functions  for both kinds of molecular junctions should differ, and the edge
induced  transmission channel should be missing in the armchair-terminated
graphene  nanocontact. As shown in Fig.\ref{f2} bottom, where the transmission
function for a CN-substituted pentacene-graphene junction with armchair-terminated
graphene nanocontacts is depicted, the edge-state-induced
transmission  channels are indeed not present for this type of molecular
junction,  whereas all of the regular peaks similar to those found in the
zigzag-terminated  graphene nanocontact appear. This result thus unambiguously
demonstrates  that these additional transmission channels are induced by graphene edge states.

\begin{figure}
  \includegraphics{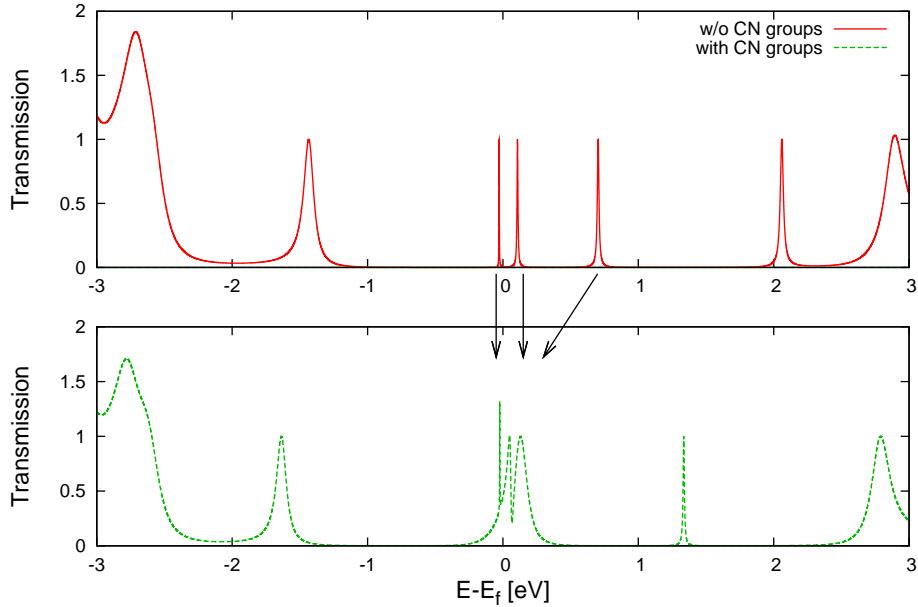}
  \caption{Transmission  through strongly coupled pentacene-graphene
    junctions. The upper and lower panel depict results without and with
    CN-substitution of the molecular bridge, respectively (see models
    d and  e in Fig.\ref{f1}). The arrows show how regular and induced
    channels  interact in the strong coupling regime to form complex peak
    structures in the transmission function.}
  \label{f4}
\end{figure}

In the molecular nanojunctions considered above, the overall coupling between
the $\pi$-electron systems of pentacene and the graphene leads is rather
small due to the twisted (almost perpendicular) orientation of the molecular
bridge with  respect to the graphene nanocontacts (see Fig.\ref{f1}b and
c). To study charge transport for stronger molecule-lead coupling, 
we have  also investigated
the molecular junctions d and f depicted in Fig.\ref{f1}. In these junctions,
the  pentacene bridge 
binds to graphene via oxygen atoms forming oxepine rings, which facilitates a
nearly  planar
structure and therefore a stronger coupling. The transmission functions (see Fig.\ref{f4})
again show a peak structure with regular peaks related to
molecular  resonances and
edge-induced transmission channels close to the Fermi level. However, the peaks are
significantly broader than those found in the amide-bound molecular junctions
due to the stronger coupling between the graphene nanocontacts and the
pentacene  bridge. Furthermore, the overlap of some of
the peaks results in values of the transmission function that are larger than unity. 
The larger coupling also induces a significantly higher
transmission amplitude of the edge-induced transmission peaks close to the Fermi
level. In the CN-substituted junction (see Fig.\ref{f4}, bottom), this is
further  enhanced by the presence of a molecular resonance close to the Fermi
level,  a result already found for the amide-bound CN-substituted
junction. Interference  between this molecular resonance and
the edge-induced structures results in a complex triple peak structure, a
characteristic  feature that can be observed in the transmission function at
the Fermi  energy (see Fig.\ref{f4}, bottom). 

In summary, we have investigated charge transport in molecule-graphene
nanojunctions.  The results obtained for different pentacene-based
nanojunctions reveal  the unique properties of graphene as material for
electrodes.  In particular, we have shown that charge transport is strongly
influenced  by the edge structure of the graphene electrodes. Graphene
electrodes with  zigzag termination have edge states that result in
additional  transport channels close to the Fermi level that strongly affect
the conductance at small bias voltages. While the influence of graphene edge
states on transport in nanojunctions has been discussed before \cite{zanolli-2010, ryndyk-2012},  the present results show, to the best of our knowledge for
the first time, the importance of edge state effects in molecule-graphene
nanojunctions in the resonant transport regime based on first principles
calculations. Our results also  show that the transport properties are
strongly  influenced by the effective  coupling between the molecular bridge
and  the graphene nanocontacts as well as the position of the molecular energy
levels,  which can be controlled  by the choice of different linker groups and appropriate
functionalization  of the pentacene molecular bridge, respectively. These findings may
facilitate  further experimental studies and possible future applications in
nanoelectronic devices.

\acknowledgement

We thank D.A.\ Ryndyk and K.\ Richter for interesting discussions. 
MT gratefully acknowledges the hospitality of the IAS at the Hebrew University Jerusalem 
within the workshop on molecular electronics. 
The generous allocation of computing time by the computing centers 
in Erlangen (RRZE), Munich (LRZ), and J\"ulich (JSC) is gratefully acknowledged. 
This work has been supported by the 
German-Israeli Foundation for Scientific Development (GIF) and the 
Deutsche Forschungsgemeinschaft (DFG) 
through  SFB 953 and the Cluster of Excellence 'Engineering of Advanced Materials'.

\bibliography{edgestates}

\providecommand*\mcitethebibliography{\thebibliography}
\csname @ifundefined\endcsname{endmcitethebibliography}
  {\let\endmcitethebibliography\endthebibliography}{}
\begin{mcitethebibliography}{40}
\providecommand*\natexlab[1]{#1}
\providecommand*\mciteSetBstSublistMode[1]{}
\providecommand*\mciteSetBstMaxWidthForm[2]{}
\providecommand*\mciteBstWouldAddEndPuncttrue
  {\def\EndOfBibitem{\unskip.}}
\providecommand*\mciteBstWouldAddEndPunctfalse
  {\let\EndOfBibitem\relax}
\providecommand*\mciteSetBstMidEndSepPunct[3]{}
\providecommand*\mciteSetBstSublistLabelBeginEnd[3]{}
\providecommand*\EndOfBibitem{}
\mciteSetBstSublistMode{f}
\mciteSetBstMaxWidthForm{subitem}{(\alph{mcitesubitemcount})}
\mciteSetBstSublistLabelBeginEnd
  {\mcitemaxwidthsubitemform\space}
  {\relax}
  {\relax}

\bibitem[Reed et~al.(1997)Reed, Zhou, Muller, Burgin, and Tour]{reed-1997}
Reed,~M.~A.; Zhou,~C.; Muller,~C.~J.; Burgin,~T.~P.; Tour,~J.~M. Conductance of
  a Molecular Junction. \emph{Science} \textbf{1997}, \emph{278},
  252--254\relax
\mciteBstWouldAddEndPuncttrue
\mciteSetBstMidEndSepPunct{\mcitedefaultmidpunct}
{\mcitedefaultendpunct}{\mcitedefaultseppunct}\relax
\EndOfBibitem
\bibitem[Park et~al.(2000)Park, Park, Lim, Anderson, Alivisatos, and
  McEuen]{park-2000}
Park,~H.; Park,~J.; Lim,~A.; Anderson,~E.; Alivisatos,~A.; McEuen,~P.
  Nanomechanical Oscillations in a Single-C60 Transistor. \emph{Nature}
  \textbf{2000}, \emph{407}, 57--60\relax
\mciteBstWouldAddEndPuncttrue
\mciteSetBstMidEndSepPunct{\mcitedefaultmidpunct}
{\mcitedefaultendpunct}{\mcitedefaultseppunct}\relax
\EndOfBibitem
\bibitem[Smit et~al.(2002)Smit, Noat, Untiedt, Lang, van Hemert, and van
  Ruitenbeek]{smit-2002}
Smit,~R.; Noat,~Y.; Untiedt,~C.; Lang,~N.; van Hemert,~M.; van Ruitenbeek,~J.
  Measurement of the Conductance of a Hydrogen Molecule. \emph{Nature}
  \textbf{2002}, \emph{419}, 906--909\relax
\mciteBstWouldAddEndPuncttrue
\mciteSetBstMidEndSepPunct{\mcitedefaultmidpunct}
{\mcitedefaultendpunct}{\mcitedefaultseppunct}\relax
\EndOfBibitem
\bibitem[Reichert et~al.(2002)Reichert, Ochs, Beckmann, Weber, Mayor, and
  L\"ohneysen]{reichert-2002}
Reichert,~J.; Ochs,~R.; Beckmann,~D.; Weber,~H.~B.; Mayor,~M.;
  L\"ohneysen,~H.~v. Driving Current through Single Organic Molecules.
  \emph{Phys. Rev. Lett.} \textbf{2002}, \emph{88}, 176804\relax
\mciteBstWouldAddEndPuncttrue
\mciteSetBstMidEndSepPunct{\mcitedefaultmidpunct}
{\mcitedefaultendpunct}{\mcitedefaultseppunct}\relax
\EndOfBibitem
\bibitem[Qiu et~al.(2004)Qiu, Nazin, and Ho]{qiu-2004}
Qiu,~X.~H.; Nazin,~G.~V.; Ho,~W. Vibronic States in Single Molecule Electron
  Transport. \emph{Phys. Rev. Lett.} \textbf{2004}, \emph{92}, 206102\relax
\mciteBstWouldAddEndPuncttrue
\mciteSetBstMidEndSepPunct{\mcitedefaultmidpunct}
{\mcitedefaultendpunct}{\mcitedefaultseppunct}\relax
\EndOfBibitem
\bibitem[Elbing et~al.(2005)Elbing, Ochs, Koentopp, Fischer, von H\"anisch,
  Weigend, Evers, Weber, and Mayor]{elbing-2005}
Elbing,~M.; Ochs,~R.; Koentopp,~M.; Fischer,~M.; von H\"anisch,~C.;
  Weigend,~F.; Evers,~F.; Weber,~H.~B.; Mayor,~M. A Single-Molecule Diode.
  \emph{Proc. Natl. Acad. Sci. U.S.A.} \textbf{2005}, \emph{102},
  8815--8820\relax
\mciteBstWouldAddEndPuncttrue
\mciteSetBstMidEndSepPunct{\mcitedefaultmidpunct}
{\mcitedefaultendpunct}{\mcitedefaultseppunct}\relax
\EndOfBibitem
\bibitem[Tao(2006)]{tao-2006}
Tao,~N. Electron Transport in Molecular Junctions. \emph{Nat. Nanotechnol.}
  \textbf{2006}, \emph{1}, 173--181\relax
\mciteBstWouldAddEndPuncttrue
\mciteSetBstMidEndSepPunct{\mcitedefaultmidpunct}
{\mcitedefaultendpunct}{\mcitedefaultseppunct}\relax
\EndOfBibitem
\bibitem[Ioffe et~al.(2008)Ioffe, Shamai, Ophir, Noy, Yutsis, Kfir,
  Cheshnovsky, and Selzer]{ioffe-2008}
Ioffe,~Z.; Shamai,~T.; Ophir,~A.; Noy,~G.; Yutsis,~I.; Kfir,~K.;
  Cheshnovsky,~O.; Selzer,~Y. Detection of Heating in Current-Carrying
  Molecular Junctions by Raman Scattering. \emph{Nat. Nanotechnol.}
  \textbf{2008}, \emph{3}, 727--732\relax
\mciteBstWouldAddEndPuncttrue
\mciteSetBstMidEndSepPunct{\mcitedefaultmidpunct}
{\mcitedefaultendpunct}{\mcitedefaultseppunct}\relax
\EndOfBibitem
\bibitem[van~der Molen and Liljeroth(2010)van~der Molen, and
  Liljeroth]{vandermolen-2010}
van~der Molen,~S.~J.; Liljeroth,~P. Charge Transport through Molecular
  Switches. \emph{J. Phys.: Condens. Matter} \textbf{2010}, \emph{22},
  133001\relax
\mciteBstWouldAddEndPuncttrue
\mciteSetBstMidEndSepPunct{\mcitedefaultmidpunct}
{\mcitedefaultendpunct}{\mcitedefaultseppunct}\relax
\EndOfBibitem
\bibitem[Ballmann et~al.(2012)Ballmann, H\"artle, Coto, Elbing, Mayor, Bryce,
  Thoss, and Weber]{ballmann-2012}
Ballmann,~S.; H\"artle,~R.; Coto,~P.~B.; Elbing,~M.; Mayor,~M.; Bryce,~M.~R.;
  Thoss,~M.; Weber,~H.~B. Experimental Evidence for Quantum Interference and
  Vibrationally Induced Decoherence in Single-Molecule Junctions. \emph{Phys.
  Rev. Lett.} \textbf{2012}, \emph{109}, 056801\relax
\mciteBstWouldAddEndPuncttrue
\mciteSetBstMidEndSepPunct{\mcitedefaultmidpunct}
{\mcitedefaultendpunct}{\mcitedefaultseppunct}\relax
\EndOfBibitem
\bibitem[H{\"a}nggi et~al.(2002)H{\"a}nggi, Ratner, and Yaliraki]{hanggi-2002}
H{\"a}nggi,~P.; Ratner,~M.; Yaliraki,~S. Processes in Molecular Wires.
  \emph{Chem. Phys.} \textbf{2002}, \emph{281}, 111--502\relax
\mciteBstWouldAddEndPuncttrue
\mciteSetBstMidEndSepPunct{\mcitedefaultmidpunct}
{\mcitedefaultendpunct}{\mcitedefaultseppunct}\relax
\EndOfBibitem
\bibitem[Nitzan and Ratner(2003)Nitzan, and Ratner]{nitzan-2003}
Nitzan,~A.; Ratner,~M.~A. Electron Transport in Molecular Wire Junctions.
  \emph{Science} \textbf{2003}, \emph{300}, 1384--1389\relax
\mciteBstWouldAddEndPuncttrue
\mciteSetBstMidEndSepPunct{\mcitedefaultmidpunct}
{\mcitedefaultendpunct}{\mcitedefaultseppunct}\relax
\EndOfBibitem
\bibitem[Cuniberti et~al.(2005)Cuniberti, Fagas, and Richter]{cuniberti}
Cuniberti,~G.; Fagas,~G.; Richter,~K. \emph{Introducing Molecular Electronics};
  Springer: Heidelberg, Germany, 2005\relax
\mciteBstWouldAddEndPuncttrue
\mciteSetBstMidEndSepPunct{\mcitedefaultmidpunct}
{\mcitedefaultendpunct}{\mcitedefaultseppunct}\relax
\EndOfBibitem
\bibitem[Joachim and Ratner(2005)Joachim, and Ratner]{joachim-2005}
Joachim,~C.; Ratner,~M.~A. Molecular Electronics: Some Views on Transport
  Junctions and Beyond. \emph{Proc. Natl. Acad. Sci. U.S.A.} \textbf{2005},
  \emph{102}, 8801--8808\relax
\mciteBstWouldAddEndPuncttrue
\mciteSetBstMidEndSepPunct{\mcitedefaultmidpunct}
{\mcitedefaultendpunct}{\mcitedefaultseppunct}\relax
\EndOfBibitem
\bibitem[Galperin et~al.(2008)Galperin, Nitzan, and Ratner]{galperin-2008}
Galperin,~M.; Nitzan,~A.; Ratner,~M.~A. The Non-Linear Response of Molecular
  Junctions: The Polaron Model Revisited. \emph{J. Phys.: Condens. Matter}
  \textbf{2008}, \emph{20}, 374107\relax
\mciteBstWouldAddEndPuncttrue
\mciteSetBstMidEndSepPunct{\mcitedefaultmidpunct}
{\mcitedefaultendpunct}{\mcitedefaultseppunct}\relax
\EndOfBibitem
\bibitem[Cuevas and Scheer(2010)Cuevas, and Scheer]{cuevas}
Cuevas,~J.; Scheer,~E. \emph{Molecular Electronics: An Introduction to Theory
  and Experiment}; World Scientific: Singapore, 2010\relax
\mciteBstWouldAddEndPuncttrue
\mciteSetBstMidEndSepPunct{\mcitedefaultmidpunct}
{\mcitedefaultendpunct}{\mcitedefaultseppunct}\relax
\EndOfBibitem
\bibitem[H\"artle and Thoss(2011)H\"artle, and Thoss]{hartle-2011}
H\"artle,~R.; Thoss,~M. Resonant Electron Transport in Single-Molecule
  Junctions: Vibrational Excitation, Rectification, Negative Differential
  Resistance, and Local Cooling. \emph{Phys. Rev. B} \textbf{2011}, \emph{83},
  115414\relax
\mciteBstWouldAddEndPuncttrue
\mciteSetBstMidEndSepPunct{\mcitedefaultmidpunct}
{\mcitedefaultendpunct}{\mcitedefaultseppunct}\relax
\EndOfBibitem
\bibitem[Feldman et~al.(2008)Feldman, Steigerwald, Guo, and
  Nuckolls]{feldman-2008}
Feldman,~A.~K.; Steigerwald,~M.~L.; Guo,~X.; Nuckolls,~C. Molecular Electronic
  Devices Based on Single-Walled Carbon Nanotube Electrodes. \emph{Acc. Chem.
  Res.} \textbf{2008}, \emph{41}, 1731--1741\relax
\mciteBstWouldAddEndPuncttrue
\mciteSetBstMidEndSepPunct{\mcitedefaultmidpunct}
{\mcitedefaultendpunct}{\mcitedefaultseppunct}\relax
\EndOfBibitem
\bibitem[Del~Valle et~al.(2007)Del~Valle, Guti{\'e}rrez, Tejedor, and
  Cuniberti]{delvalle-2007}
Del~Valle,~M.; Guti{\'e}rrez,~R.; Tejedor,~C.; Cuniberti,~G. Tuning the
  Conductance of a Molecular Switch. \emph{Nat. Nanotechnol.} \textbf{2007},
  \emph{2}, 176--179\relax
\mciteBstWouldAddEndPuncttrue
\mciteSetBstMidEndSepPunct{\mcitedefaultmidpunct}
{\mcitedefaultendpunct}{\mcitedefaultseppunct}\relax
\EndOfBibitem
\bibitem[Castro~Neto et~al.(2009)Castro~Neto, Guinea, Peres, Novoselov, and
  Geim]{castroneto-2009}
Castro~Neto,~A.~H.; Guinea,~F.; Peres,~N. M.~R.; Novoselov,~K.~S.; Geim,~A.~K.
  The Electronic Properties of Graphene. \emph{Rev. Mod. Phys.} \textbf{2009},
  \emph{81}, 109--162\relax
\mciteBstWouldAddEndPuncttrue
\mciteSetBstMidEndSepPunct{\mcitedefaultmidpunct}
{\mcitedefaultendpunct}{\mcitedefaultseppunct}\relax
\EndOfBibitem
\bibitem[Liu et~al.(2009)Liu, Ryu, Chen, Steigerwald, Nuckolls, and
  Brus]{liu-2009}
Liu,~H.; Ryu,~S.; Chen,~Z.; Steigerwald,~M.~L.; Nuckolls,~C.; Brus,~L.~E.
  Photochemical Reactivity of Graphene. \emph{J. Am. Chem. Soc.} \textbf{2009},
  \emph{131}, 17099--17101\relax
\mciteBstWouldAddEndPuncttrue
\mciteSetBstMidEndSepPunct{\mcitedefaultmidpunct}
{\mcitedefaultendpunct}{\mcitedefaultseppunct}\relax
\EndOfBibitem
\bibitem[Cao et~al.(2009)Cao, Liu, Shen, Yan, Li, Xu, Yu, Steigerwald,
  Nuckolls, Liu, and {et al.}]{cao-2009}
Cao,~Y.; Liu,~S.; Shen,~Q.; Yan,~K.; Li,~P.; Xu,~J.; Yu,~D.;
  Steigerwald,~M.~L.; Nuckolls,~C.; Liu,~Z.; {et al.}, High-Performance
  Photoresponsive Organic Nanotransistors with Single-Layer Graphenes as
  Two-Dimensional Electrodes. \emph{Adv. Func. Mat.} \textbf{2009}, \emph{19},
  2743--2748\relax
\mciteBstWouldAddEndPuncttrue
\mciteSetBstMidEndSepPunct{\mcitedefaultmidpunct}
{\mcitedefaultendpunct}{\mcitedefaultseppunct}\relax
\EndOfBibitem
\bibitem[Prins et~al.(2011)Prins, Barreiro, Ruitenberg, Seldenthuis,
  Aliaga-Alcalde, Vandersypen, and van~der Zant]{prins-2011}
Prins,~F.; Barreiro,~A.; Ruitenberg,~J.~W.; Seldenthuis,~J.~S.;
  Aliaga-Alcalde,~N.; Vandersypen,~L. M.~K.; van~der Zant,~H. S.~J.
  Room-Temperature Gating of Molecular Junctions Using Few-Layer Graphene
  Nanogap Electrodes. \emph{Nano Lett.} \textbf{2011}, \emph{11},
  4607--4611\relax
\mciteBstWouldAddEndPuncttrue
\mciteSetBstMidEndSepPunct{\mcitedefaultmidpunct}
{\mcitedefaultendpunct}{\mcitedefaultseppunct}\relax
\EndOfBibitem
\bibitem[Jin et~al.(2009)Jin, Lan, Peng, Suenaga, and Iijima]{jin-2009}
Jin,~C.; Lan,~H.; Peng,~L.; Suenaga,~K.; Iijima,~S. Deriving Carbon Atomic
  Chains from Graphene. \emph{Phys. Rev. Lett.} \textbf{2009}, \emph{102},
  205501\relax
\mciteBstWouldAddEndPuncttrue
\mciteSetBstMidEndSepPunct{\mcitedefaultmidpunct}
{\mcitedefaultendpunct}{\mcitedefaultseppunct}\relax
\EndOfBibitem
\bibitem[Bergvall et~al.(2011)Bergvall, Berland, Hyldgaard, Kubatkin, and
  L\"ofwander]{bergvall-2011}
Bergvall,~A.; Berland,~K.; Hyldgaard,~P.; Kubatkin,~S.; L\"ofwander,~T.
  Graphene Nanogap for Gate-Tunable Quantum-Coherent Single-Molecule
  Electronics. \emph{Phys. Rev. B} \textbf{2011}, \emph{84}, 155451\relax
\mciteBstWouldAddEndPuncttrue
\mciteSetBstMidEndSepPunct{\mcitedefaultmidpunct}
{\mcitedefaultendpunct}{\mcitedefaultseppunct}\relax
\EndOfBibitem
\bibitem[Ryndyk et~al.(2012)Ryndyk, Bundesmann, Liu, and Richter]{ryndyk-2012}
Ryndyk,~D.~A.; Bundesmann,~J.; Liu,~M.-H.; Richter,~K. Edge State Effects in
  Junctions with Graphene Electrodes. \emph{Phys. Rev. B} \textbf{2012},
  \emph{86}, 195425\relax
\mciteBstWouldAddEndPuncttrue
\mciteSetBstMidEndSepPunct{\mcitedefaultmidpunct}
{\mcitedefaultendpunct}{\mcitedefaultseppunct}\relax
\EndOfBibitem
\bibitem[Zanolli et~al.(2010)Zanolli, Onida, and Charlier]{zanolli-2010}
Zanolli,~Z.; Onida,~G.; Charlier,~J.-C. Quantum Spin Transport in Carbon
  Chains. \emph{ACS Nano} \textbf{2010}, \emph{4}, 5174--5180\relax
\mciteBstWouldAddEndPuncttrue
\mciteSetBstMidEndSepPunct{\mcitedefaultmidpunct}
{\mcitedefaultendpunct}{\mcitedefaultseppunct}\relax
\EndOfBibitem
\bibitem[Forrest(2004)]{forrest-2004}
Forrest,~S. The Path to Ubiquitous and Low-Cost Organic Electronic Appliances
  on Plastic. \emph{Nature} \textbf{2004}, \emph{428}, 911--918\relax
\mciteBstWouldAddEndPuncttrue
\mciteSetBstMidEndSepPunct{\mcitedefaultmidpunct}
{\mcitedefaultendpunct}{\mcitedefaultseppunct}\relax
\EndOfBibitem
\bibitem[Katz and Huang(2009)Katz, and Huang]{katz-2009}
Katz,~H.~E.; Huang,~J. Thin-Film Organic Electronic Devices. \emph{Annu. Rev.
  Mat. Res.} \textbf{2009}, \emph{39}, 71--92\relax
\mciteBstWouldAddEndPuncttrue
\mciteSetBstMidEndSepPunct{\mcitedefaultmidpunct}
{\mcitedefaultendpunct}{\mcitedefaultseppunct}\relax
\EndOfBibitem
\bibitem[Frisch et~al.()Frisch, Trucks, Schlegel, Scuseria, Robb, Cheeseman,
  Scalmani, Barone, Mennucci, Petersson, and {et al.}]{G09}
Frisch,~M.~J.; Trucks,~G.~W.; Schlegel,~H.~B.; Scuseria,~G.~E.; Robb,~M.~A.;
  Cheeseman,~J.~R.; Scalmani,~G.; Barone,~V.; Mennucci,~B.; Petersson,~G.~A.;
  {et al.}, Gaussian~09 {R}evision {C}.1. Gaussian Inc. Wallingford CT
  2009\relax
\mciteBstWouldAddEndPuncttrue
\mciteSetBstMidEndSepPunct{\mcitedefaultmidpunct}
{\mcitedefaultendpunct}{\mcitedefaultseppunct}\relax
\EndOfBibitem
\bibitem[TUR()]{TURBOMOLE}
{TURBOMOLE V6.3 2011}, a development of {University of Karlsruhe} and
  {Forschungszentrum Karlsruhe GmbH}, 1989-2007, {TURBOMOLE GmbH}, since 2007;
  available from {\tt http://www.turbomole.com}.\relax
\mciteBstWouldAddEndPunctfalse
\mciteSetBstMidEndSepPunct{\mcitedefaultmidpunct}
{}{\mcitedefaultseppunct}\relax
\EndOfBibitem
\bibitem[Benesch et~al.(2008)Benesch, {\v{C}}{\'\i}{\v{z}}ek, Klime{\v{s}},
  Kondov, Thoss, and Domcke]{benesch-2008}
Benesch,~C.; {\v{C}}{\'\i}{\v{z}}ek,~M.; Klime{\v{s}},~J.; Kondov,~I.;
  Thoss,~M.; Domcke,~W. Vibronic Effects in Single Molecule Conductance:
  First-Principles Description and Application to Benzenealkanethiolates
  between Gold Electrodes. \emph{J. Phys. Chem. C} \textbf{2008}, \emph{112},
  9880--9890\relax
\mciteBstWouldAddEndPuncttrue
\mciteSetBstMidEndSepPunct{\mcitedefaultmidpunct}
{\mcitedefaultendpunct}{\mcitedefaultseppunct}\relax
\EndOfBibitem
\bibitem[Kopf and Saalfrank(2004)Kopf, and Saalfrank]{kopf-2004}
Kopf,~A.; Saalfrank,~P. Electron Transport through Molecules Treated by LCAO-MO
  Green's Functions with Absorbing Boundaries. \emph{Chem. Phys. Lett.}
  \textbf{2004}, \emph{386}, 17 -- 24\relax
\mciteBstWouldAddEndPuncttrue
\mciteSetBstMidEndSepPunct{\mcitedefaultmidpunct}
{\mcitedefaultendpunct}{\mcitedefaultseppunct}\relax
\EndOfBibitem
\bibitem[Datta(1997)]{datta}
Datta,~S. \emph{Electronic Transport in Mesoscopic Systems}; Cambridge
  University Press, 1997\relax
\mciteBstWouldAddEndPuncttrue
\mciteSetBstMidEndSepPunct{\mcitedefaultmidpunct}
{\mcitedefaultendpunct}{\mcitedefaultseppunct}\relax
\EndOfBibitem
\bibitem[Gutierrez et~al.(2003)Gutierrez, Fagas, Richter, Grossmann, and
  Schmidt]{gutierrez-2003}
Gutierrez,~R.; Fagas,~G.; Richter,~K.; Grossmann,~F.; Schmidt,~R. Conductance
  of a Molecular Junction Mediated by Unconventional Metal-Induced Gap States.
  \emph{Europhys. Lett.} \textbf{2003}, \emph{62}, 90--96\relax
\mciteBstWouldAddEndPuncttrue
\mciteSetBstMidEndSepPunct{\mcitedefaultmidpunct}
{\mcitedefaultendpunct}{\mcitedefaultseppunct}\relax
\EndOfBibitem
\bibitem[Fagas et~al.(2004)Fagas, Gutierrez, Richter, Grossmann, and
  Schmidt]{fagas-2004}
Fagas,~G.; Gutierrez,~R.; Richter,~K.; Grossmann,~F.; Schmidt,~R. Manifestation
  of Electrode Surface States in Molecular Conduction. \emph{Macromol. Symp.}
  \textbf{2004}, \emph{212}, 103--112\relax
\mciteBstWouldAddEndPuncttrue
\mciteSetBstMidEndSepPunct{\mcitedefaultmidpunct}
{\mcitedefaultendpunct}{\mcitedefaultseppunct}\relax
\EndOfBibitem
\bibitem[Benesch et~al.(2006)Benesch, {\v{C}}{\'\i}{\v{z}}ek, Thoss, and
  Domcke]{benesch-2006}
Benesch,~C.; {\v{C}}{\'\i}{\v{z}}ek,~M.; Thoss,~M.; Domcke,~W. Vibronic Effects
  on Resonant Electron Conduction through Single Molecule Junctions.
  \emph{Chem. Phys. Lett.} \textbf{2006}, \emph{430}, 355--360\relax
\mciteBstWouldAddEndPuncttrue
\mciteSetBstMidEndSepPunct{\mcitedefaultmidpunct}
{\mcitedefaultendpunct}{\mcitedefaultseppunct}\relax
\EndOfBibitem
\bibitem[Xue and Ratner(2003)Xue, and Ratner]{xue-2003}
Xue,~Y.; Ratner,~M.~A. Microscopic Study of Electrical Transport through
  Individual Molecules with Metallic Contacts. I. Band Lineup, Voltage Drop,
  and High-Field Transport. \emph{Phys. Rev. B} \textbf{2003}, \emph{68},
  115406\relax
\mciteBstWouldAddEndPuncttrue
\mciteSetBstMidEndSepPunct{\mcitedefaultmidpunct}
{\mcitedefaultendpunct}{\mcitedefaultseppunct}\relax
\EndOfBibitem
\bibitem[Wakabayashi et~al.(2009)Wakabayashi, Takane, Yamamoto, and
  Sigrist]{wakabayashi-2009}
Wakabayashi,~K.; Takane,~Y.; Yamamoto,~M.; Sigrist,~M. Electronic Transport
  Properties of Graphene Nanoribbons. \emph{New J. Phys.} \textbf{2009},
  \emph{11}, 095016\relax
\mciteBstWouldAddEndPuncttrue
\mciteSetBstMidEndSepPunct{\mcitedefaultmidpunct}
{\mcitedefaultendpunct}{\mcitedefaultseppunct}\relax
\EndOfBibitem
\end{mcitethebibliography}


\end{document}